\documentclass[12pt, a4paper]{article}
\usepackage{authblk}

\newcommand{\avg}[1]{\langle #1 \rangle}

\newcommand{\heading}[1]{\noindent \textbf{#1}}

\usepackage{graphicx}
\usepackage{cases}
\usepackage[breaklinks=true,colorlinks=true,linkcolor=blue,urlcolor=blue,citecolor=blue]{hyperref}
\usepackage[margin=1in]{geometry}
\usepackage{type1cm}
\usepackage{lettrine}
\usepackage{graphicx}
\usepackage{epstopdf}
\usepackage{booktabs}

\begin{document}

\title{Understanding the spatiotemporal pattern of grazing cattle movement}

\date{}

\author{Kun Zhao\footnote{To whom correspondence should be addressed. E-mail: \url{kun.zhao@csiro.au}.}~}
\author{Raja Jurdak}

\affil{CSIRO Data61, Brisbane, Queensland, Australia}

\maketitle

\begin{abstract}
In this study, we analyse a high-frequency movement dataset for a group of grazing cattle and investigate their spatiotemporal patterns using a simple two-state `stop-and-move' mobility model. We find that the dispersal kernel in the moving state is best described by a mixture exponential distribution, indicating the hierarchical nature of the movement. On the other hand, the waiting time appears to be scale-invariant below a certain cut-off and is best described by a truncated power-law distribution, suggesting heterogenous dynamics in the non-moving state. We explore possible explanations for the observed phenomena, covering factors that can  play a role in the generation of mobility patterns, such as the context of grazing environment, the intrinsic decision-making mechanism or the energy status of different activities. In particular, we propose a new hypothesis that the underlying movement pattern can be attributed to the most probable observable energy status under the maximum entropy configuration. These results are not only valuable for modelling cattle movement but also provide new insights for understanding the underlying biological basis of grazing behaviour.

\end{abstract}

\newpage

\section*{Introduction}

Animal movement is a highly complex process driven by various random and deterministic mechanisms involving a large number of causing factors~\cite{Nathan2008, Kays2015}. It has been proposed that spatiotemporal patterns in movement may arise from moving strategies that evolve to optimise foraging efficiency~\cite{Viswanathan1999, Zhao2015}, decision-making processes in response to external stimuli~\cite{Wearmouth2014}, environmental conditions or landscape features~\cite{Bronikowski1996, Humphries2010, DeJager2011}, collective dynamics and social interactions~\cite{Pelletier2004, Strandburg2015}, memory and home-return behaviour~\cite{Song2010}, just to name a few. Fully unravelling the complexity of animal movement as well as sorting out the intricate relations between the observed spatiotemporal pattern and various underlying causing factors remains a difficult scientific challenge. 
 
For over a century, our attempts to understand animal movement have been limited to a qualitative level due to the lack of high-quality data that can provide fine-grained spatiotemporal description of movement~\cite{Kays2015}. Recently, new tracking technologies such as the Global Positioning System (GPS) have been deployed in animal tracking to obtain continuous time-resolved moving trajectories with high spatiotemporal resolution. The emerging high-quality movement data enables the application of quantitative analysis and mathematical characterisation on mobility patterns at different spatiotemporal scales. 

One common approach to analyse animal movement is to represent the time-resolved trajectory as discrete moving steps under the framework of random walk~\cite{Bartumeus2005, Codling2008, Reynolds2009}. In this context, the dispersal kernel $p(r)$ in space, which characterises the general distribution of step length $r$ in the trajectory, is considered to be a significant footprint of movement~\cite{Bullock2002, Brockmann2006}. The detailed functional form of $p(r)$ is indicative of a specific type of random walk and the underlying dynamics of movement. For example, an exponential kernel function $p(r) \sim e^{-r/r_c}$ with $r_c$ being the characteristic length scale is the signature of the classic Brownian walk that obeys the central-limit theorem and exhibits a normal-diffusive pattern. A scaling dispersal kernel characterised by a power-law function $p(r) \sim r^{-\gamma}$ with $\gamma$ being the scaling exponent is the signature of the L\'{e}vy walk which exhibits high heterogeneity and super-diffusive pattern. Much effort has been devoted to study the dispersal kernel $p(r)$ for different animal species using real movement data, from small insects like honey bees~\cite{Reynolds2007}, marine life like jelly fish and whales~\cite{Sims2008}, birds like albatross~\cite{Viswanathan1996, Humphries2012}, to mammals like monkeys~\cite{RamosFG2004} and human~\cite{Brockmann2006, Gonzalez2008, Raichlen2014}. One controversial topic that attracts tremendous attention in this line is about whether the observed movement follows a Brownian-like motion or a L\'{e}vy walk. Although many studies have suggested strong evidence for the existence of L\'{e}vy walks in animals, it has been argued that this evidence may come from statistical artefacts or inappropriate manipulation of data~\cite{Edwards2007, Benhamou2007, Clauset2009}. Indeed, the dispersal kernel $p(r)$ only provides partial information on the spatial pattern and does not fully capture all important movement aspects such as the temporal spectrum that depicts the switch among different activity modes over time. Recently it has been found that scaling phenomena in movement can also arise in the waiting time distribution $p(\tau_w)$ that characterises the time span of non-moving period, or the inter-event time distribution $p(\tau_e)$ that characterises the time  between two successive moving activities~\cite{Brockmann2006, Song2010, Rhee2011}. These findings suggest there is a need for more detailed investigation on the spatiotemporal pattern in movement beyond the dispersal kernel.        

\begin{figure}[htb!]
\centering
\includegraphics[width=6in]{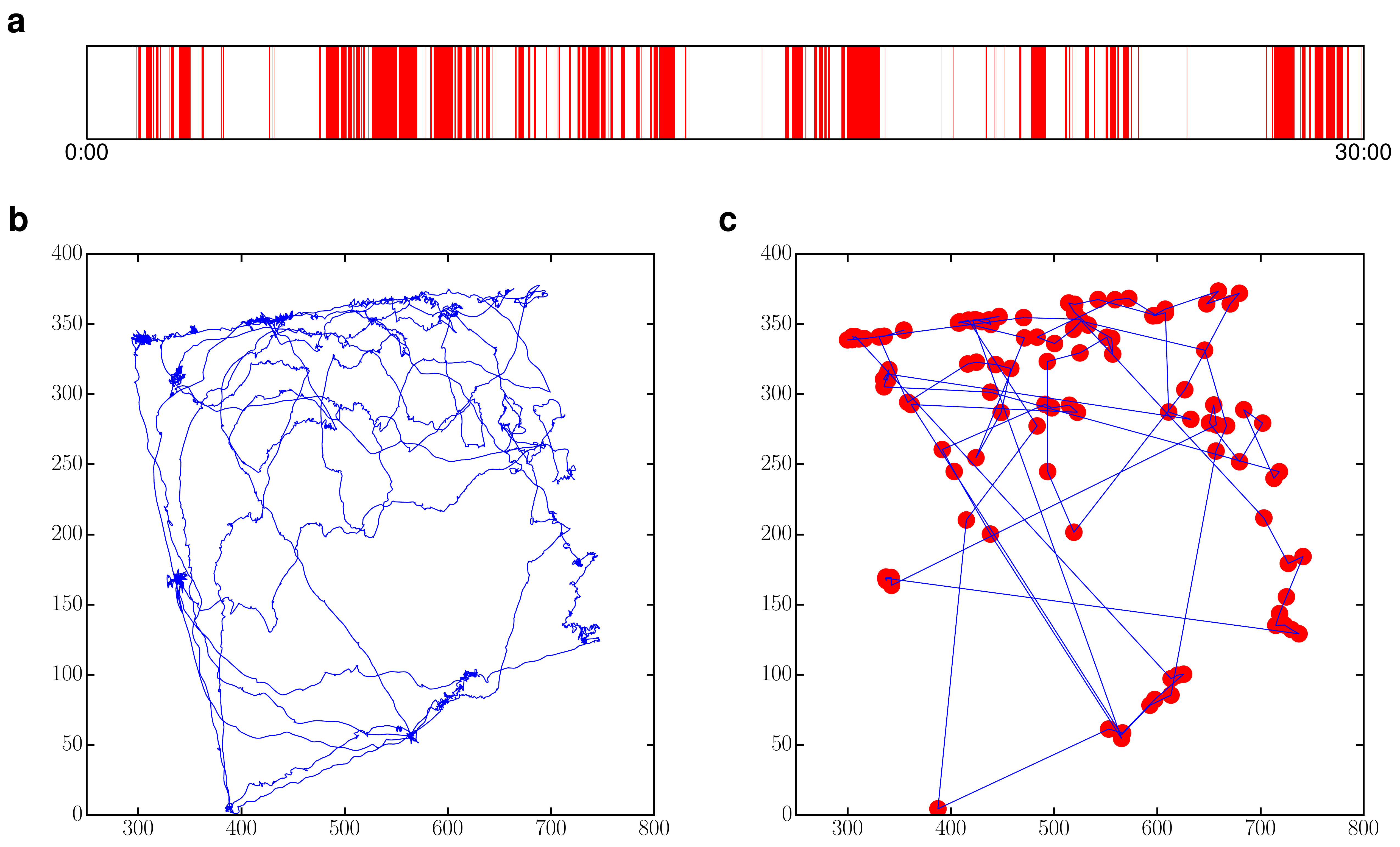}
\caption{A schematic diagram of the spatiotemporal pattern under the two-state `stop-and-move' representation. (a) The temporal spectrum of activities illustrated in a spike train. Colour segments on the time-axis represent alternating waiting (white) and moving (red) activities over time. (b) The raw trajectory of an individual cow before processing. (c) The spatial pattern extracted from the raw trajectory in panel (b) can be projected as a transition graph,  where the waiting locations for non-moving segments are represented by red dots and the trips are represented by blue solid lines. }
\label{fig:scheme}
\end{figure}

Here we use a dataset of high-frequency GPS samples to study the movement of grazing cattle. The continuous time-resolved trajectories allow us to gain  comprehensive insight into the spatiotemporal pattern. We use a two-state `stop-and-move' model to describe the mobility pattern, dividing the trajectory into alternate moving and non-moving states (see \textit{Material and Methods}). Specifically, the non-moving state indicates that the animal remains within a radius $\Delta r$ in space for at least $\Delta t$ in time, where $\Delta r$ represents the spatial resolution limit in the observation and $\Delta t$ is a tuning threshold parameter to specify the minimum time span. The non-moving segment in the trajectory can be viewed as a single point in space, which we call waiting location, with a length $\tau_w \ge \Delta t$ in time, which we call waiting time correspondingly.  On the other hand, the moving state indicates that the animal is in a transition from one waiting location to another, which can be described as a trip $(l, \tau_m)$ in the trajectory with $l$ being the distance between the two waiting locations and $\tau_m$ being the time elapse of the trip. The representation of mobility pattern in this approach is shown in the schematic diagram of Fig.~\ref{fig:scheme}. 

Under this representation, we observe a very interesting spatiotemporal pattern in which the two activity states are of unique statistical characterization. In particular, we find that the dispersal kernel or trip length distribution $p(l)$ is best described by a hybrid exponential distribution, which indicates that the trajectory has a two-level hierarchical structure in space and each level appears to follow a Brownian walk. This is in contrast to the widely-observed L\'{e}vy walk patterns in other species. Despite the absence of scaling law in the spatial dispersal, we find that the waiting distribution $p(\tau_w)$ in the time domain is best described by a truncated power-law. Possible underlying mechanisms and ecological implications accounting for this phenomena are discussed (see \textit{Discussions}). Our results provide new quantitative insights into grazing cattle movement that are lacking in most previous work. Indeed, understanding grazing/foraging animal movements is not only a critical issue in biological science but also of fundamental importance to many practical issues such as farm and livestock management~\cite{Amy2001}, the maintenance of biodiversity in ecosystems~\cite{Daszak2000} and developing better tracking~\cite{Sommer16} and virtual fencing technologies~\cite{Jurdak10}.

\section*{Results}

\begin{figure}[htb!]
\centering
\includegraphics[width=6in]{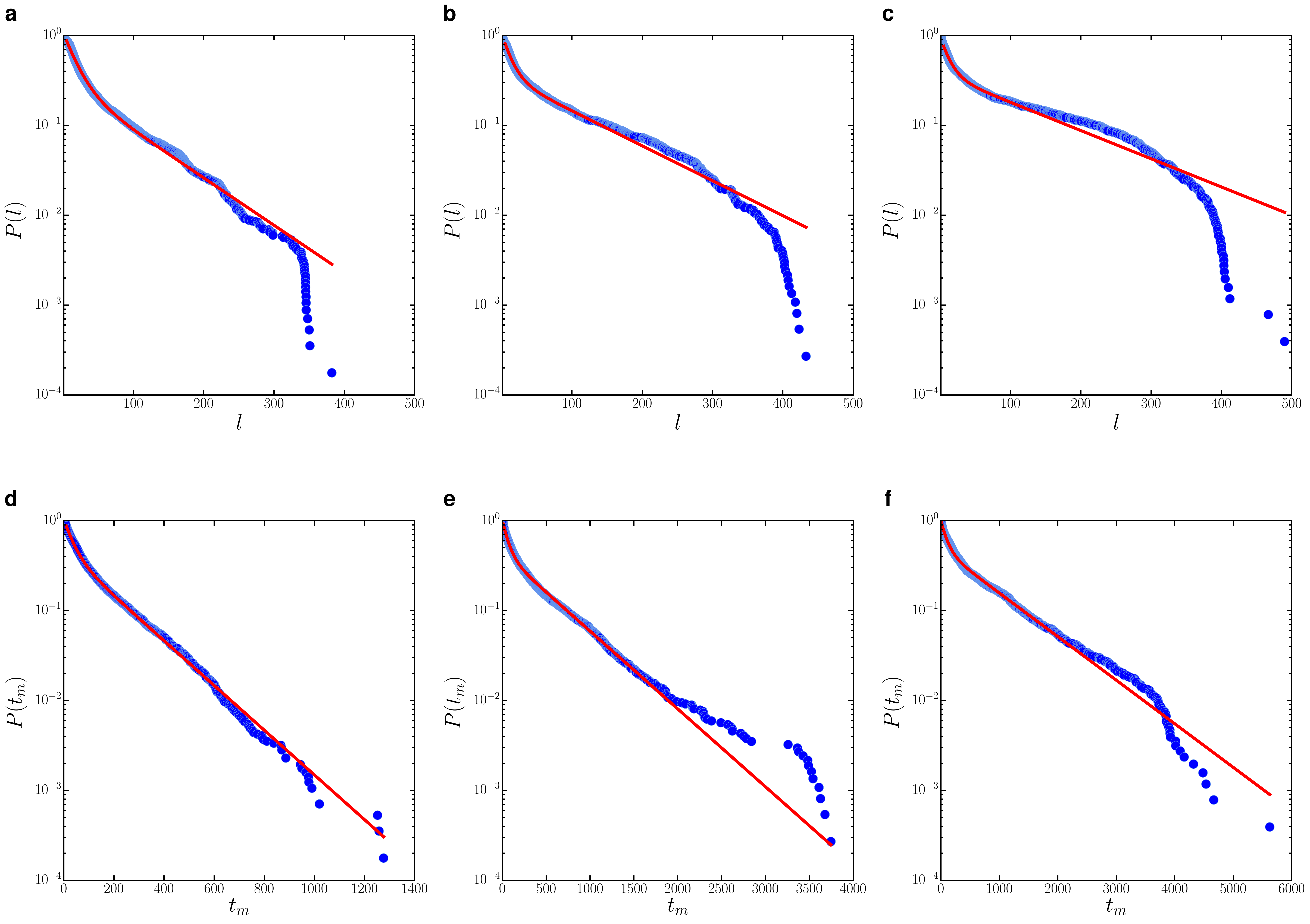}
\caption{The statistics for the moving state with $\Delta r = 5$m. Panels (a) - (c) are the cumulative distributions $P(l)$ for trip length with $\Delta t = 1, 2, 3$~mins (from left to right). Panels (d) - (f) are the cumulative distributions $P(
\tau_m)$ for trip time with $\Delta t = 1, 2, 3$~mins (from left to right). The solid red lines represent the best fitted mixture exponential obtained by the maximum-likelihood method using an expectation-maximisation algorithm.}
\label{fig:spatial}
\end{figure}

\heading{Dispersal kernel in moving state.} We first turn attention to the spatial dispersal kernel or the trip length distribution $p(l)$ over the whole population. Obtaining the functional form of $p(l)$ directly from empirical data requires a binning process, which has been known to have statistical distortion for data with a broad distribution~\cite{Clauset2009}. To avoid the disadvantage of binning, we use the complementary cumulative distribution $P(l) \equiv \int_l p(l)dl$ for statistical analysis. We process the data using three different parameter sets with $\Delta r = 5$m corresponding to the resolution limit of positioning device and $\Delta t = 1,2,3$~mins respectively~\cite{Godsk2011}. To describe the dispersal kernel $P(l)$ shown in Fig.~\ref{fig:spatial} a-c, we consider four commonly-used candidate models~\cite{Jansen2012}: (1) power-law; (2) truncated power-law with exponential cut-off; (3) exponential; (4) mixture exponential. Using the maximum-likelihood estimation (MLE) to fit the candidate models and the Akaike-information-criterion (AIC) for model selection~\cite{Edwards2007} (see \textit{Supporting Information}), we find that the best model to describe $P(l)$ is the mixture exponential  
\begin{equation}
P(l) = qe^{-(l-l_{min})/l_1} + (1-q)e^{-(l-l_{min})/l_2},
\label{eq:mixexp2}
\end{equation}
where $l_1$ and $l_2$ are the characteristic lengths in each mixture component, $q$ is a parameter specifying the mixture proportion, and $l_{min} = \Delta r = 5$m is the lower bound in observation.  Another significant statistical feature of the moving state is the trip time distribution $p(\tau_m)$, which is also best described by the mixture exponential model, as shown in Fig.~\ref{fig:spatial} d-f. This is consistent with our expectation that the trip time $\tau_m$ is strongly correlated with the trip length $l$. 

\begin{figure}[htb!]
\centering
\includegraphics[width=6in]{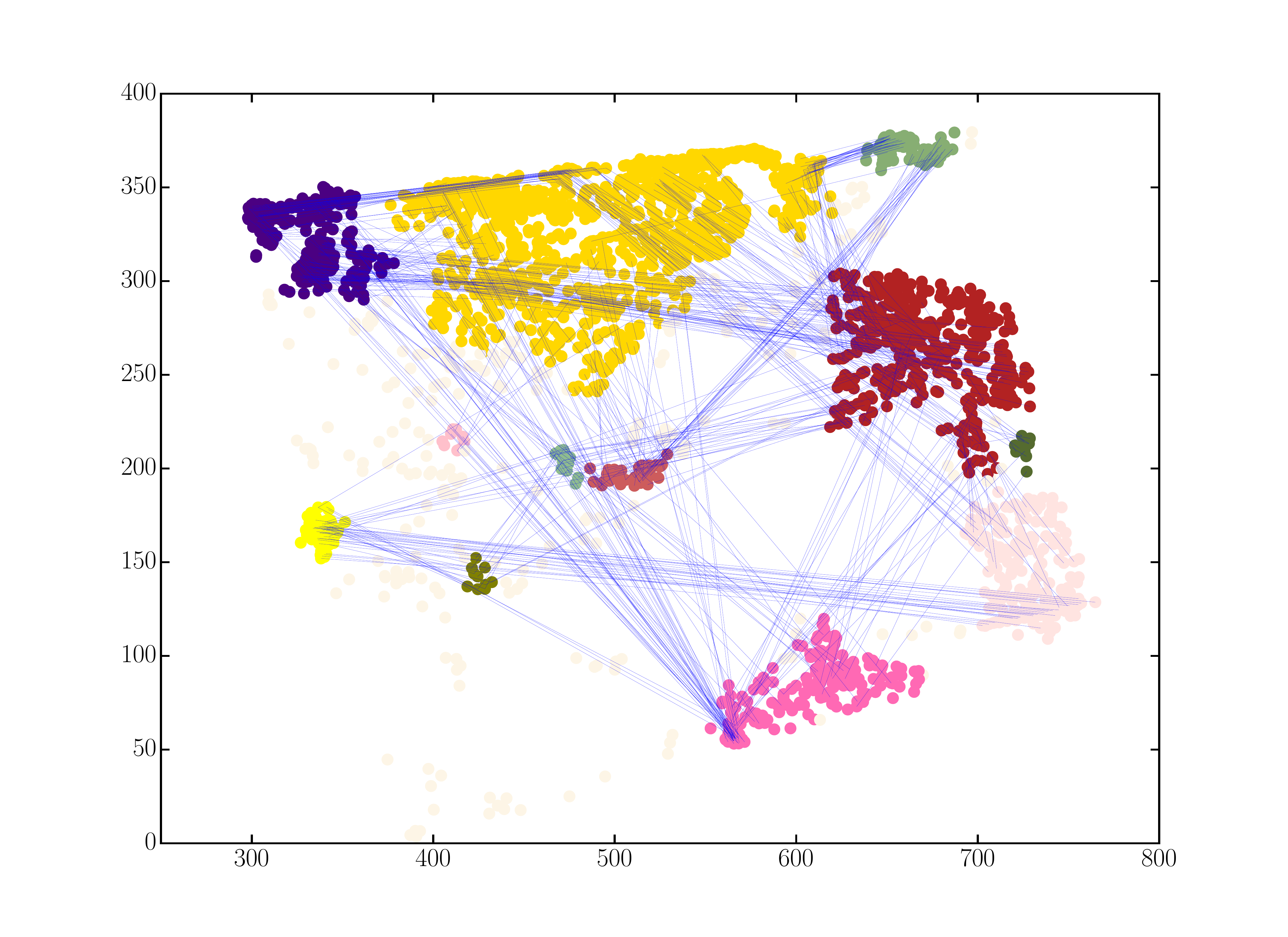}
\caption{The visualisation of clusters extracted by the DBSCAN algorithms and the corresponding inter-cluster trips. Dots with different colours represent different clusters (the lightest colour represent outliers). Blue solid lines indicate inter-cluster trips. }
\label{fig:cluster}
\end{figure}

\begin{figure}[htb!]
\centering
\includegraphics[width=6in]{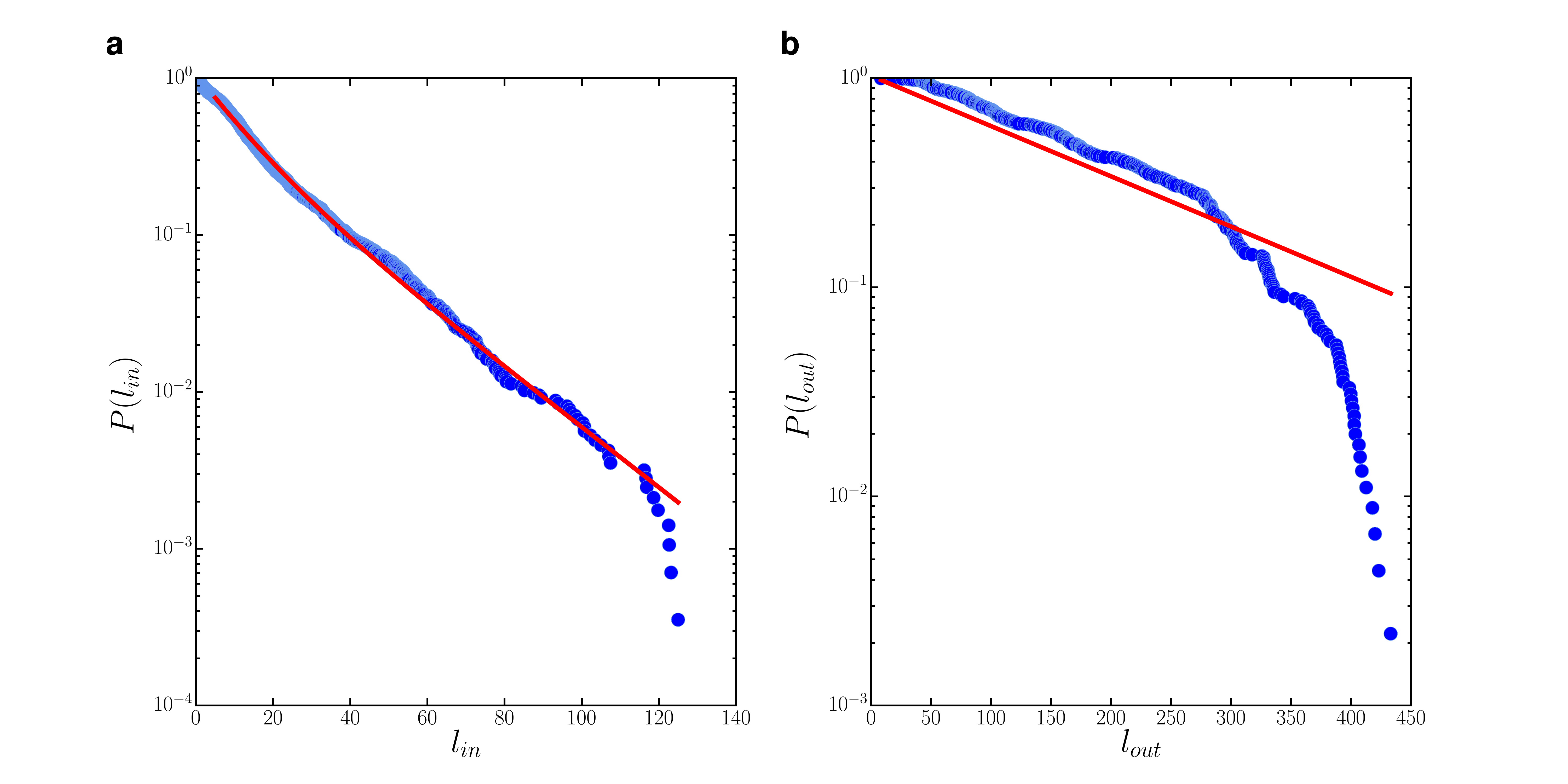}
\caption{The trip length distribution for intra-cluster movements and inter-cluster movements. Both of them are well described by an exponential distribution (red solid lines). }
\label{fig:cluster_stat}
\end{figure}

The mixture exponential here indicates that the spatial pattern of grazing cattle is governed by two different Brownian-like dynamics with different characteristic scales, suggesting a hierarchical structure of the movement. If we consider that the landscape is formed by a number of patchy areas,  the first exponential distribution will represent the short-range movement that occurs within a patch, and the second exponential distribution with a larger characteristic length will represent inter-patch movements. To better reveal this hierarchy structure in mobility pattern, we perform clustering on the waiting locations using a density-based clustering algorithm DBSCAN which is efficient in discovering significant clusters with irregular shape from noisy data points, as shown in Fig.~\ref{fig:cluster}. After grouping the waiting locations into clusters, the trips in the mobility pattern fall into two categories, intra-cluster trip and inter-cluster trip. We find that the trip length distribution for each of these two types of trips can be well described by a single exponential distribution, as shown in Fig.~\ref{fig:cluster_stat}. It is worth noting that the mixture exponential still renders the highest AIC weight among the four candidate models for the intra-cluster trip length distribution, while the single exponential has the highest AIC weight without the mixture exponential. However the difference between the two components in the mixture model is comparably small ($l_1 = 11.20, l_2 = 22.05, q = 0.44, \Delta t= 2$~mins), suggesting that the two components are not strongly distinguishable and a single exponential is a reasonable alternative model in this scenario.  We also observe that some long-distance inter-cluster trips are of high similarity, indicating that transitions from one cluster to another are not completely random and spontaneous, but could be driven by a deterministic process such as memory or herding.  \\

\begin{figure}[htb!]
\centering
\includegraphics[width=6in]{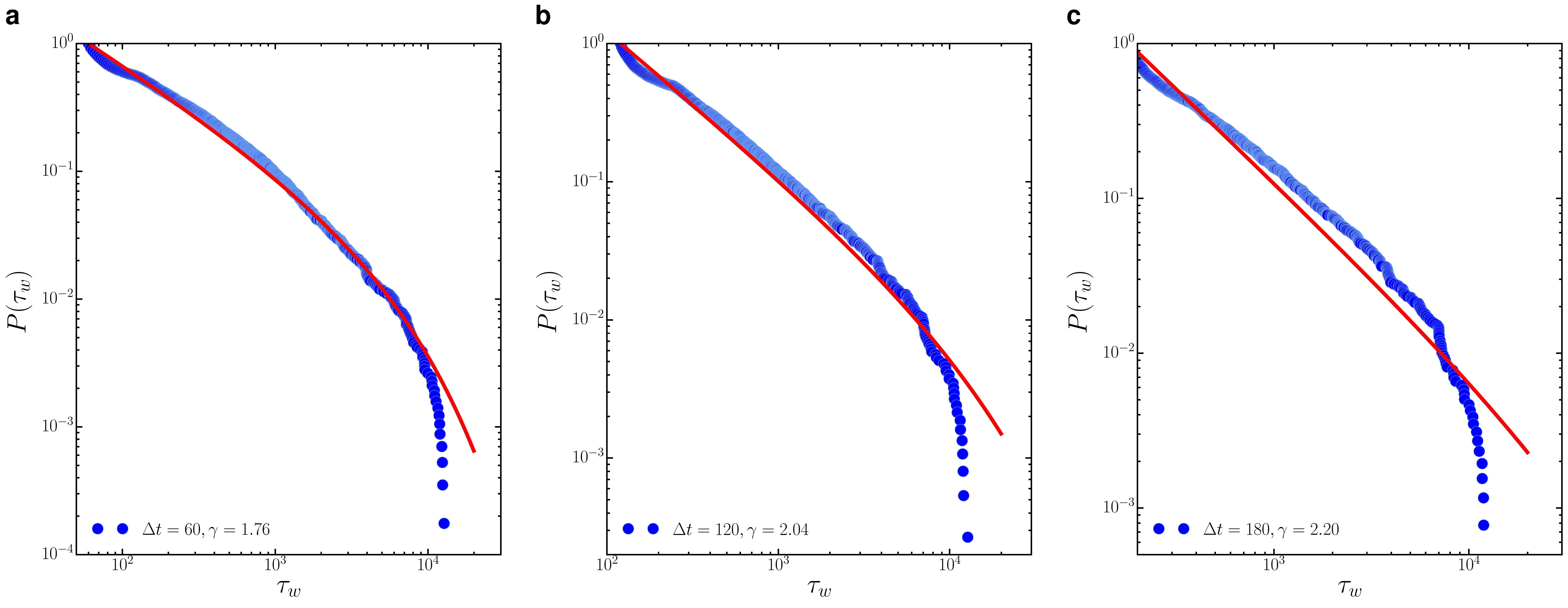}
\caption{The statistics for the moving state with $\Delta r = 5$m. Panels (a) - (c) are the cumulative distributions $P(\tau_w)$ for waiting time with $\Delta t = 1, 2, 3$~mins (from left to right). The solid red lines represent the best fitted truncated power-law obtained by the maximum-likelihood method. }
\label{fig:waiting}
\end{figure}

\heading{Waiting time distribution.}  To characterise the waiting time distribution, we compare three different models, namely exponential, power-law and truncated power-law with exponential cut-off (see \textit{Supporting Information}). We observe that the waiting-time distribution $p(\tau_w)$ is best described by a truncated power-law distribution 
\begin{equation}
p(\tau_w) \sim \tau_w ^{-\gamma} e^{-\tau_w/ \tau_w^c}
\end{equation}
with $\gamma$ being the scaling exponent and $\tau_w^c$ being the structural cut-off. As shown in Fig.~\ref{fig:waiting}, varying the threshold parameter $\Delta t$ in data processing does not affect the emergence of scaling phenomena. The scaling law in the waiting time distribution is indicative of the heterogeneous grazing dynamics of cattle, which could be related to the landscape heterogeneity,  the complex decision-making dynamics or the energy management of movement (See \textit{Discussion}). We also find that the waiting time distributions in the main clusters discovered by the DBSCAN algorithm are all well described by a truncated power-law, suggesting the scaling behaviour in waiting time distribution is invariant at the cluster-level.   

Without taking into account correlation between activities, the temporal spectrum of mobility pattern can be approximated as a two-state renewal process where the time span of the alternate moving and non-moving activities are randomly drawn from the distribution functions $p(\tau_m)$ and $p(\tau_w)$ respectively. To test the validity of this approximation, we measure the pairwise Pearson correlation coefficient between the time span of consecutive activity segments in the following four situations: (1) the non-moving segment and the next moving segment ($r = -0.0344, p = 0.0363$); (2) the moving segment and the next non-moving segment ($r = -0.0616, p = 0.000174$); (3) two consecutive non-moving segment ($r = 0.0787, p = 1.58 \times 10^{-6}$); (4) two consecutive moving segment ($r = 0.0719, p = 1.24 \times 10^{-6}$). We find that none of these shows significant correlation. The result suggests that short-range correlation does not exist in the temporal spectrum, i.e. the time span of the previous activity has little influence on the time span of the next activity, and the temporal dynamics can be approximately described by a two-state renewal process without considering long-range correlation. \\

\heading{Individual mobility pattern.} 
The population-based statistics presented above are not necessarily representative of the individual patterns. It has been suggested that the characteristics of population statistics may differ from their individual counterpart after being aggregated over population. For example, the observed L\'{e}vy walk pattern in population may arise from individual heterogeneity~\cite{Petrovskii2011}.  To test whether the individual pattern is consistent with the population-based statistics, we use the same model selection procedure to fit the individual statistics ($\Delta r = 5m$, $\Delta t = 2$~mins). We find that the trip length and trip time distribution for each individual is best described by the hybrid exponential, with only one exception in the trip length distribution. On the other hand, the waiting distribution for each individual is best described by power-law or truncated power-law (see \textit{Supporting Information}). This suggests that the composite Brownian walk in space as well as the scaling law in waiting time distribution are not an statistical artefact due to the mixture of different individual patterns, but they appear to be universal for all individuals.  Although all individual spatiotemporal patterns are best described by the same distribution functions, the fitted parameters vary from individual to individual. For example, the exponents of the truncated power-law for waiting distribution estimated by the maximum-likelihood method range from $\gamma = 1.6$ to $\gamma = 2.5$. This indicates that the internal properties encapsulated by the scaling exponent $\gamma$ are different among individual cows, although their activities appear to be governed by the same dynamics. \\

\section*{Discussion}
In this study we have found that under a two-state `stop-and-move' representation the spatiotemporal pattern of grazing cattle exhibits a hierarchical structure in space and an asymmetric temporal spectrum, which can be described by a composite Brownian walk interspersed with power-law distributed non-moving periods. This finding is in contrast to the patterns observed in human~\cite{Rhee2011} and T-cell~\cite{Harris2012} mobility where the moving and non-moving states are both characterised by a scaling law. Since detailed statistical characterisation on free-range animal movement based on high-frequency GPS trajectories is still largely missing, this finding can provide new perspectives to our understanding for grazing animals movement and useful leads to the underlying ecological basis of grazing behaviour.  

A simple deterministic scenario that can give rise to the observed scaling law in waiting time distribution is that the environment is structured according to the same heterogenous statistics. We can consider that different location $\vec{r}$ in the landscape is of different quality or resource abundance described a quality function $Q(\vec{r})$. If the cattle simply spend their time for feeding on one location proportional to the quality $Q(\vec{r})$ at that location, i.e. a `greedy' strategy, $Q(\vec{r})$ would be the observed waiting distribution. 

Stochastic processes and spontaneous behaviour can also account for the observed spatiotemporal pattern. Recently, a plausible decision-based queueing process in which the animal executes activities from a stochastic priority list has been used to interpret the scaling law observed in the waiting time of marine predators~\cite{Sims2008}. This model was originally proposed to explain the power-law distributed inter-event time observed in the communication pattern in human dynamics~\cite{Gonzalez2008}. Specifically, the model assumes that the animal performs the two activities waiting and moving with probability $x_1$ and $x_2 = 1 - x_1$ at a regular basis, where $x_1$ and $x_2$ are the priority of the activity drawn from a random distribution $p(x)$. If the animal moves, it changes its context and therefore its likelihood to move or stay also changes. As a result, the priority will be redrawn from the random distribution, representing the change of state due to the movement. This model can generate the power-law distribution in waiting time as well as the exponential distribution in step-size. By introducing a deterministic component to the decision probability, the model can be also tuned to generate different scaling exponent $\gamma$ accounting for the various scaling phenomena in different species. The model is recast in a dynamic prey-predator environment where the moving probability $x_1$ can be interpreted as the likelihood of finding a prey in the vicinity. 

We can also consider the movement as a two-state point process, in which the probabilities that the animal switches its state are $q_A$ (from moving to non-moving) and $q_B$ (from non-moving to moving)~\cite{Karsai2012, Ginelli2015}. It is well known that the state duration is exponential distributed when the switching probability is constant and independent of time~\cite{Ginelli2015}. Recently, it has been suggested that the power-law distributed duration can be attributed to the reinforcement dynamics, such that the switching probability is proportional to the time that animal has spend in its current state, i.e. the longer the animal stays in its current state the less likely it will change it~\cite{Zhao2011a, Zhao2011b, Karsai2012}.    

Another explanation is to associate the movement pattern with the energy state of the animal using a maximum entropy approach. In this context, each moving and non-moving activity is associated with a certain amount of energy loss $E_l$ or energy gain $E_g$. According to the maximum entropy principle, the distribution of $E_l$ and $E_g$ over all activity segments should follow a Boltzmann distribution $p(E_{l,g}) \sim e^{-E_{l,g}}$ (See \textit{Material and methods}). The validation of the maximum entropy approach is mainly subject to two conditions: (1) each individual activity is independent and has no influence on others; and (2) the energy intake and expenditure is maintained by two different mechanisms and can be treated as two isolated systems. The first condition is supported by our test on the correlation between consecutive activities, while the second is intuitively understandable. Following this formulation, it is straightforward to derive that when $E_g \propto \log{\tau_w}$ and $E_l \propto \tau_m$ the observed scaling law in waiting time as well as the exponential distribution in trip time can be reproduced. That is to say, the energy intake increases logarithmically as grazing time increases, while the energy expenditure due to moving increases linearly with the moving time or distance. It is interesting to note that the logarithmic energy intake function has been suggested for grazing animals before, and the linear energy expenditure or cost function has been widely observed in many single-mode movements of human transportation activities~\cite{Kolbl2003, Yan2013}. It is well known that energy status can affect animal movement, but a quantitative understanding of their relation is still unclear. Our proposed maximum entropy approach can potentially fill this gap by establishing a connection between the energy function and the observed mobility pattern, suggesting that the detailed energy intake or expenditure as a function of time in different activities can be inferred from statistical features of the macroscopic mobility patterns such as waiting time or step-length distributions. The conjectured relation can be tested in future experiments by measuring detailed energy intake or consumption using laboratory techniques.

\section*{Material and methods}

\heading{Dataset description.}
The dataset consists of continuous $0.5$-Hz GPS samples for $34$ individuals covering an observation period of over $50$ hours. We select the data of $31$ individuals in which there is no discontinuity in GPS samples and we choose a continuous 30-hour observation window during which the animals were grazing in a confined 600m $\times$ 400m rectangular area. The trajectory for each individual cow can be denoted by a sequence $L = \{p_i\}$, where $p_i = (x_i, y_i, t_i)$ represents a GPS sample with $(x_i, y_i)$ being the position coordinates and $t_i$ being the timestamp. We use moving average filtering to reduce the noise and smooth the trajectory with a $10$ sec moving window, such that $p_i = \avg{p_{i-2}, p_{i-1}, p_i, p_{i+1}, p_{i+2}}$. \\

\heading{Classification of mobility pattern.}
We define the non-moving segment of a trajectory as a set of consecutive points $L_w = \{p_k, p_{k+1},\dots,p_{k+m-1}\}$, which satisfies the following three conditions: (1) the distance $d_{k,j}$ from the starting point $p_k$ to any other point $p_j$ of the segment must be smaller than a certain threshold $\Delta d$, i.e. $\max_{k<j<k+m}{d_{k,j}} \le \Delta r$; (2) the distance from the starting point $p_k$ to the point following the ending point of the segment $p_{k+m}$ must be larger than $\Delta r$, i.e. $d_{k,k+m}> \Delta r$; (3) the time span of the segment must be longer than a certain threshold $\Delta t$, i.e. $t_{k+m} - t_{k} > \Delta t$. In this definition, the first two constraints are made to identify consecutive points that are likely to represent an identical position within in a certain proximity. The third constraint imposes a minimum time span of the non-moving segment that can be tuned to exclude some very-short random activities such as a pause when encountering an obstacle, as well as making the extracted non-moving segments more representative of meaningful activities such as grazing or resting. After extracting the non-moving segments, we simply define the points between two non-moving segments as the moving segments. The approach here is in analogy to the definition of staying points for continuous GPS samples in most spatiotemporal analysis of human mobility~\cite{Rhee2011, Jiang2013}. The value of $\Delta t$ is suggested to be $2-3$ mins~\cite{Godsk2011}. \\ 


\heading{Maximum entropy principle.} 
The maximum entropy principle originates from statistical mechanics, which assumes that the configuration of microscopic states of a complex system (e.g. the energy of each particle) leading to the macroscopic observation is the one that maximise the entropy of the system. Suppose the system consists of $N$ non-interacting particles and has a total energy $U$, such that $N = \sum n_i$ and $U = \sum n_iE_i$ where $n_i$ denotes the number of particles at a specific energy state $E_i$. Then the ensemble that represents all possible configurations of the system is called the canonical ensemble, and the probability $p(E)$ that a particle has a specific energy state $E$ is denoted by $P(E) \propto e^{-E/\bar{E}}$. Here we assume that the alternate moving and non-moving activities in the mobility pattern operate in two independent systems, while the individual activities are regarded as `particles' and the associated energy state of the activity is the incurred energy gain (or loss) due to the activity. To obtain the distribution of the time span $\tau$ in each activity,  we use the transformation $p(\tau) = p(E) \frac{dE(\tau)}{d\tau}$ where $E(\tau)$ is the energy function that describes the energy gain as a function of the time span during the activity. The detailed form of the distribution function $p(\tau)$ is then subject to the energy function $E(\tau)$. For example, a logarithmic function $E(\tau) \propto \log{\tau}$ will lead to a power-law distribution $p(\tau) \propto \tau^{\beta}$, while a linear function will simply maintain a exponential form $p(\tau) \propto e^{-k\tau}$.

\section*{Acknowledgements}

\newpage{\pagestyle{empty}\cleardoublepage}

\clearpage

\section*{Supporting Information}
\begin{table*}[ht]
\centering
\caption{The AIC weights for the population trip length distribution}
\begin{tabular}{ccccc}
$\Delta t$  &  Powerlaw  &  Truncated powerlaw  &  Exponential  &  Hybrid exponential\\ \hline 
60  &  0.00  &  0.00  &  0.00  &  \textbf{1.00} \\  
120  &  0.00  &  0.00  &  0.00  &  \textbf{1.00} \\  
180  &  0.00  &  0.00  &  0.00  &  \textbf{1.00} \\  
\hline 
\end{tabular}

\end{table*}

\begin{table*}[ht]
\centering
\caption{The AIC weights for the population trip time distribution}
\begin{tabular}{ccccc}
$\Delta t$  &  Powerlaw  &  Truncated powerlaw  &  Exponential  &  Hybrid exponential\\ \hline 
60  &  0.00  &  0.00  &  0.00  &  \textbf{1.00} \\  
120  &  0.00  &  0.00  &  0.00  &  \textbf{1.00} \\  
180  &  0.00  &  0.00  &  0.00  &  \textbf{1.00} \\  
\hline 
\end{tabular}

\end{table*}

\begin{table*}[ht]
\centering
\caption{The AIC weights for the population waiting time distribution}
\begin{tabular}{cccc}
$\Delta t$  &  Powerlaw  &  Truncated powerlaw  &  Exponential\\ \hline 
60  &  0.00  &  \textbf{1.00}&  0.00 \\  
120  &  0.00  &  \textbf{1.00}&  0.00 \\  
180  &  \textbf{0.52}&  0.48  &  0.00 \\  
\hline 
\end{tabular}

\end{table*}

\begin{table*}[ht]
\centering
\caption{The AIC weights for the individual trip length distribution}
\begin{tabular}{ccccc}
Cow ID  &  Powerlaw  &  Truncated powerlaw  &  Exponential  &  Hybrid exponential\\ \hline 
0  &  0.00  &  0.40  &  0.00  &  \textbf{0.60} \\  
1  &  0.00  &  0.01  &  0.00  &  \textbf{0.99} \\  
2  &  0.00  &  0.06  &  0.00  &  \textbf{0.94} \\  
4  &  0.00  &  0.35  &  0.00  &  \textbf{0.65} \\  
5  &  0.00  &  0.08  &  0.00  &  \textbf{0.92} \\  
6  &  0.00  &  0.23  &  0.00  &  \textbf{0.77} \\  
7  &  0.00  &  0.05  &  0.00  &  \textbf{0.95} \\  
8  &  0.00  &  0.14  &  0.00  &  \textbf{0.86} \\  
9  &  0.00  &  0.01  &  0.00  &  \textbf{0.99} \\  
10  &  0.00  &  0.11  &  0.00  &  \textbf{0.89} \\  
12  &  0.00  &  0.47  &  0.00  &  \textbf{0.53} \\  
13  &  0.00  &  0.14  &  0.00  &  \textbf{0.86} \\  
14  &  0.00  &  0.01  &  0.00  &  \textbf{0.99} \\  
16  &  0.00  &  0.40  &  0.00  &  \textbf{0.60} \\  
17  &  0.00  &  0.22  &  0.00  &  \textbf{0.78} \\  
18  &  0.00  &  0.02  &  0.00  &  \textbf{0.98} \\  
19  &  0.00  &  0.05  &  0.00  &  \textbf{0.95} \\  
20  &  0.00  &  0.03  &  0.00  &  \textbf{0.97} \\  
21  &  0.00  &  0.01  &  0.00  &  \textbf{0.99} \\  
22  &  0.00  &  \textbf{0.56}&  0.00  &  0.44 \\  
23  &  0.00  &  0.33  &  0.00  &  \textbf{0.67} \\  
24  &  0.00  &  0.22  &  0.00  &  \textbf{0.78} \\  
25  &  0.00  &  0.00  &  0.00  &  \textbf{1.00} \\  
26  &  0.00  &  0.02  &  0.00  &  \textbf{0.98} \\  
27  &  0.00  &  0.01  &  0.00  &  \textbf{0.99} \\  
29  &  0.00  &  0.17  &  0.00  &  \textbf{0.83} \\  
30  &  0.00  &  0.45  &  0.00  &  \textbf{0.55} \\  
32  &  0.00  &  0.27  &  0.00  &  \textbf{0.73} \\  
33  &  0.00  &  0.22  &  0.00  &  \textbf{0.78} \\  
\hline 
\end{tabular}

\end{table*}

\begin{table*}[ht]
\centering
\caption{The AIC weights for the individual trip time distribution}
\begin{tabular}{ccccc}
Cow ID  &  Powerlaw  &  Truncated powerlaw  &  Exponential  &  Hybrid exponential\\ \hline 
0  &  0.00  &  0.00  &  0.00  &  \textbf{1.00} \\  
1  &  0.00  &  0.00  &  0.00  &  \textbf{1.00} \\  
2  &  0.00  &  0.00  &  0.00  &  \textbf{1.00} \\  
4  &  0.00  &  0.00  &  0.00  &  \textbf{1.00} \\  
5  &  0.00  &  0.00  &  0.00  &  \textbf{1.00} \\  
6  &  0.00  &  0.00  &  0.00  &  \textbf{1.00} \\  
7  &  0.00  &  0.00  &  0.00  &  \textbf{1.00} \\  
8  &  0.00  &  0.00  &  0.00  &  \textbf{1.00} \\  
9  &  0.00  &  0.00  &  0.00  &  \textbf{1.00} \\  
10  &  0.00  &  0.00  &  0.00  &  \textbf{1.00} \\  
12  &  0.00  &  0.00  &  0.00  &  \textbf{1.00} \\  
13  &  0.00  &  0.00  &  0.00  &  \textbf{1.00} \\  
14  &  0.00  &  0.00  &  0.00  &  \textbf{1.00} \\  
16  &  0.00  &  0.00  &  0.00  &  \textbf{1.00} \\  
17  &  0.00  &  0.00  &  0.00  &  \textbf{1.00} \\  
18  &  0.00  &  0.00  &  0.00  &  \textbf{1.00} \\  
19  &  0.00  &  0.00  &  0.00  &  \textbf{1.00} \\  
20  &  0.00  &  0.00  &  0.00  &  \textbf{1.00} \\  
21  &  0.00  &  0.00  &  0.00  &  \textbf{1.00} \\  
22  &  0.00  &  0.00  &  0.00  &  \textbf{1.00} \\  
23  &  0.00  &  0.00  &  0.00  &  \textbf{1.00} \\  
24  &  0.00  &  0.00  &  0.00  &  \textbf{1.00} \\  
25  &  0.00  &  0.00  &  0.00  &  \textbf{1.00} \\  
26  &  0.00  &  0.00  &  0.00  &  \textbf{1.00} \\  
27  &  0.00  &  0.00  &  0.00  &  \textbf{1.00} \\  
29  &  0.00  &  0.00  &  0.00  &  \textbf{1.00} \\  
30  &  0.00  &  0.00  &  0.00  &  \textbf{1.00} \\  
32  &  0.00  &  0.00  &  0.00  &  \textbf{1.00} \\  
33  &  0.00  &  0.00  &  0.00  &  \textbf{1.00} \\  
\hline 
\end{tabular}

\end{table*}

\begin{table*}[ht]
\centering
\caption{The AIC weights for the individual waiting time distribution}
\begin{tabular}{cccc}
Cow ID  &  Powerlaw  &  Truncated powerlaw  &  Exponential\\ \hline 
0  &  \textbf{0.66}&  0.34  &  0.00 \\  
1  &  \textbf{0.66}&  0.34  &  0.00 \\  
2  &  \textbf{0.71}&  0.29  &  0.00 \\  
4  &  \textbf{0.72}&  0.28  &  0.00 \\  
5  &  \textbf{0.68}&  0.32  &  0.00 \\  
6  &  \textbf{0.66}&  0.34  &  0.00 \\  
7  &  0.49  &  \textbf{0.51}&  0.00 \\  
8  &  \textbf{0.73}&  0.27  &  0.00 \\  
9  &  \textbf{0.73}&  0.27  &  0.00 \\  
10  &  \textbf{0.66}&  0.34  &  0.00 \\  
12  &  \textbf{0.62}&  0.38  &  0.00 \\  
13  &  \textbf{0.58}&  0.42  &  0.00 \\  
14  &  \textbf{0.66}&  0.34  &  0.00 \\  
16  &  \textbf{0.73}&  0.27  &  0.00 \\  
17  &  \textbf{0.73}&  0.27  &  0.00 \\  
18  &  \textbf{0.72}&  0.28  &  0.00 \\  
19  &  0.48  &  \textbf{0.52}&  0.00 \\  
20  &  0.48  &  \textbf{0.52}&  0.00 \\  
21  &  \textbf{0.71}&  0.29  &  0.00 \\  
22  &  0.36  &  \textbf{0.64}&  0.00 \\  
23  &  0.47  &  \textbf{0.53}&  0.00 \\  
24  &  \textbf{0.61}&  0.39  &  0.00 \\  
25  &  \textbf{0.70}&  0.30  &  0.00 \\  
26  &  \textbf{0.70}&  0.30  &  0.00 \\  
27  &  \textbf{0.71}&  0.29  &  0.00 \\  
29  &  \textbf{0.68}&  0.32  &  0.00 \\  
30  &  0.49  &  \textbf{0.51}&  0.00 \\  
32  &  \textbf{0.62}&  0.38  &  0.00 \\  
33  &  \textbf{0.62}&  0.38  &  0.00 \\  
\hline 
\end{tabular}

\end{table*}

\end{document}